\documentclass{aa}  
\usepackage{natbib}
\bibpunct{(}{)}{;}{a}{}{,} 
\usepackage{graphicx}
\usepackage{txfonts}
\usepackage{xcolor}
\usepackage{amsmath}   
\usepackage{amssymb}   
\usepackage{mathtools}
\usepackage{multicol} 
\usepackage{multirow} 
\usepackage{threeparttable}
\usepackage{longtable}
\usepackage{tabto}
\usepackage{adjustbox}
\usepackage{comment}
\usepackage[colorlinks = true,
            linkcolor = blue,
            urlcolor  = blue,
            citecolor = blue,
            anchorcolor = blue]{hyperref}
\usepackage{lscape}

\begin{document}

   \title{New orbital periods of high-inclination dwarf novae based on\\ \textit{Gaia} Alerts photometry}


   \author{Sáez-Carvajal, C.
          \inst{1},
          \
          Vogt, N.
          \inst{1},
          \
          Zorotovic, M.
          \inst{1},
          García-Veas, J.
          \inst{1},
          \
          Aravena-Rojas, G.
          \inst{1},
          \
          Dumond, L.
          \inst{1},
          \
          \\
          Figueroa-Tapia, F.
          \inst{1},
          López-Bonilla, Y.
          \inst{1,2},
          \
          Rodriguez-Jimenez, A.
          \inst{1,3},
          Vega-Manubens, I.
          \inst{1},
          \
          Grawe, B.
          \inst{1}
          }

   \institute{Instituto de Física y Astronomía, Universidad de Valparaíso,
              Av. Gran Bretaña 1111, 5030 Casilla, Valparaíso, Chile\\
              \email{catalina.saezc@alumnos.uv.cl}
                    \and
        Instituto de Física, Pontificia Universidad Católica de Valparaíso,
              Av. Brasil 2950, 4059 Casilla, Valparaíso, Chile
              \and
        Departamento de Astronomía, Universidad de Chile,
        Casilla 36-D, Santiago, Chile}

   \date{Received 25 July 2024; accepted 28 September 2024}

 
 \abstract{The orbital period of a cataclysmic variable stands as a crucial parameter for investigating the structure and physics of these binary systems, as well as understanding their evolution. We use photometric \textit{Gaia} data for dwarf novae (DNe) in the quiescent state ---which are available for a number of years--- to determine new orbital periods and improve or modify previously suggested period values. Two approaches are implemented for selecting high-inclination targets, either eclipsing or with ellipsoidal variations. 
  We determine new orbital periods for 75 DNe and improve ephemerides for 27 more (three of which change significantly),  contributing 9.4\% of the known DNe periods of between 0.05 and 2.0 days, and doubling the number of known periods exceeding 0.44 days. Their phase-folded light curves are presented and arranged by orbital period, illustrating the transition from short-period systems, dominated by radiation from the accretion disc and the hot spot, to longer-period DNe, where the Roche-lobe-filling secondary star is the primary visual flux source. This transition ---which occurs around the well-known period gap (between $\sim$ 2 and 3 hours)--- is expected, as DNe with larger orbital periods typically harbour more massive donors, which contribute to the visible flux. However, this transition is not abrupt. Within the same range of periods, we observe systems dominated by ellipsoidal variations, where the companion star is clearly visible, as well as others dominated by the disc and the hot spot. The presence of some DNe with ellipsoidal variations near the lower edge of the period gap is striking, as the companions in these systems are expected to be cool low-mass M-dwarfs not visible in the light curve. This could indicate that we are observing systems where the donor star was originally much more massive and underwent significant nuclear evolution before mass-transfer began, as has been suggested previously for QZ Ser.}

   \keywords{stars: dwarf novae, cataclysmic variables -- methods: data analysis}

    \titlerunning{New orbital periods of high-inclination DNe based on \textit{Gaia} Alerts photometry}
        \authorrunning{Sáez-Carvajal et al.}
   \maketitle

%

\section{Introduction}

Cataclysmic variables (CVs) are short-period binary systems consisting of a white dwarf steadily accreting from a low-mass companion, typically a main sequence star (see \citealt{warner95} for a comprehensive review). Among all observational parameters that can be determined for a CV, the orbital period is by far the one that can be derived with the highest precision.
It is also a crucial parameter, as it is directly linked to the size of the binary, as the secondary stellar component of a CV always fills its Roche lobe. 

An intriguing observational feature of CVs is the notable dearth of systems with periods of between $\sim$ 2 and 3 hours, known as the `period gap'. The disrupted magnetic braking model for angular momentum loss (AML) in CVs was postulated by \citet{rappaportetal83} to explain the existence of the gap. CVs evolve towards shorter periods due to AML caused by the emission of gravitational waves \citep{paczynski+sienkiewicz81} and wind magnetic braking \citep{verbunt+zwaan81}. The disrupted magnetic braking model assumes that magnetic braking dominates over gravitational radiation when the donor star retains a radiative core and becomes significantly less efficient once the donor becomes fully convective. This transition typically occurs in main sequence stars with masses of $\sim$ 0.3\,M$_{\odot}$, corresponding to an orbital period of $\sim$ 3 hours for a Roche-Lobe-filling main sequence star in a binary system with a white dwarf primary. Therefore, CVs above this period have larger AML rates, leading to higher mass-transfer rates and donors that are out of thermal equilibrium.  The strong reduction in the efficiency of magnetic braking at $\sim$ 3 hours results in a steep decline in the AML along with the mass-transfer rate, allowing the donor star to contract towards its equilibrium radius, temporarily halting mass transfer until it refills its Roche lobe at an orbital period of $\sim$ 2 hours.
Observational support for a strong reduction of the AML rate at the fully convective boundary comes from the mass-transfer rates derived for the observed CV population \citep{Pala2022} as well as the observed inflation of donors above the period gap \citep{knigge06,kniggeetal11}. However, recent studies have raised doubts about the validity of the disrupted magnetic braking model, particularly in light of substantial contradictions between the model's predictions and observations \citep[e.g.][]{Schreiber24,Sarkar24,Schaefer24}. These studies suggest that the model's original formulation may not fully account for all observed phenomena in CVs, and alternative models are being explored.

CVs often exhibit brightness variations due to the mass transfer from the companion star onto the surface of the white dwarf. This mass transfer usually occurs via an accretion disc around the white dwarf and can lead to sudden and dramatic increases in brightness, the so-called dwarf nova (DN) outbursts, with typical amplitudes of between 4 and 7 magnitudes. These DN outbursts are caused by a thermal instability in the accretion disc, which leads to a sudden increase in the rate of accretion, resulting in a temporary brightening of the system \citep{Osaki74,Cannizoetal88}. DN outbursts typically occur at semi-regular intervals, ranging from days to a few years, and are observed in DNe within the entire orbital period range of CVs (between  $\sim$1.3 hours and $\sim$3 days).

Eclipsing CVs are particularly useful because their time-resolved photometry, combined with spectroscopy, allows us to derive the orbital inclination. This enables the determination of all the physical and geometrical parameters of the binary system. The same is true for CVs exhibiting ellipsoidal variations in their light curves. This phenomenon can be observed in systems with high inclination angles and low mass-transfer rates, where the tidally distorted donor star dominates the overall flux \citep[e.g.][]{SzkodyMateo1986}.  

Here we present two effective methods for selecting targets with high orbital inclination (either eclipsing or ellipsoidal systems) among the more than 6\,000 DNe observed by the \textit{Gaia} space telescope \citep{gaia-collab} over the past decade. Generally, \textit{Gaia}'s photometric errors are of the order of a few percent, even down to a $g$-band magnitude of $\sim$\,20. In many cases, this allows us to obtain reliable orbital periods and phased light curves from \textit{Gaia} data of DNe during their quiescent state, even though these data often have sparse and irregular time distributions, including large gaps of months or years.

\section{Period search methods applied to photometric \textit{Gaia} Alerts data}
 
We focussed our research on the \textit{Gaia} Alerts database, specifically selecting DNe from The International Variable Star Index (VSX) of the American Association of Variable Star Observers \citep[AAVSO;][]{sas25thsymposium}. This source allows us to download the light curve in the $g$-band, which covers the wavelength range from 330 to 1050 nm, providing insight into the star's brightness and variability in this specific part of the spectrum. The utility of \textit{Gaia} photometry for determining orbital periods was demonstrated in a previous study of VW Hyi \citep{saez}. This non-eclipsing DN exhibits an orbital hump, the ephemeris of which was significantly improved using photometric \textit{Gaia} data. In general, we limited our search to systems with a $g$-band magnitude of brighter than $21$ and only analysed cases with more than about 40 observations in quiescence, after eliminating any DN outburst observations through visual inspection of the original light curves.

 \subsection{Application of the Lomb-Scargle periodogram }

We used the Lomb-Scargle (LS) periodogram \citep{Lomb,Scargle} to determine a first approximation of the orbital periods. Despite the uneven distribution and occasional gaps in the data, for some targets we were able to identify a dominant peak in the periodogram indicative of a periodic signal. Within a small range of test periods around this LS period value, we displayed many phase-folded light curves using very small steps from one test period to the next, and determined by visual inspection the accurate period value at which the data give the best convergence of the phased light curve. Subsequently, we selected observations around the brightness minimum phase before proceeding with the linear fit as described in Section 2.3.

Typically, the LS analysis generated sinusoidal phased light curves characterised by a single minimum and a single maximum. However, in systems with longer periods that show ellipsoidal variations due to the deformed secondary star, we observe two maxima and two minima in each orbital revolution. In such cases, we also analysed the light curve by doubling the LS period value and examining differences in the depth of the two minima. The deepest minimum was chosen to present phase 0 in the ephemeris of the system.


\subsection{Periods determined from the eclipse times}

When inspecting the $g$-band \textit{Gaia} light curve provided by the VSX catalogue, some of the systems revealed sudden, isolated   much fainter  measurements indicative of deep eclipses. 
In those cases, all intervals between those eclipse epochs must correspond to integer multiples of the orbital period.

Following a procedure similar to that used by \citet{Schaefer2019} for QZ Vir (see their Section 5.1), we developed a code that uses a sequence of possible orbital periods to perform a systematic search in the range between 1.5 and 24 hours. 
For each potential period ($P_{\text{orb}}$), we divided all the time intervals between eclipses ($\Delta T_{ij}$) by $P_{\text{orb}}$ to determine the number of cycles ($\Delta N_{ij}$) between two observed eclipses:
\begin{equation} 
\Delta N_{ij} = \frac{\Delta T_{ij}}{P_{\text{orb}}},
\end{equation}
\noindent where $i$ and $j$ were integers corresponding to the number of the observed eclipse. A correct orbital period will result in $\Delta N_{ij}$ being an integer. To identify the most likely orbital period, we calculated the absolute differences between $\Delta N_{ij}$ and the nearest integer for each pair of suspected eclipses. We then summed these absolute differences across all pairs of eclipses. The orbital period that minimised this sum was selected as the most likely value for $P_{\text{orb}}$. The correct solution was always confirmed by a phase-folded light curve.


\subsection{Determination of the ephemeris}

The heliocentric Julian dates (HJDs) of the selected data points surrounding the phases of the eclipses were collected and the cycle count system ---denoted E ---was established, normally assigning a value of E = 0 to the first eclipse epoch observed in the data, denoted HJDo. We used a linear least-square fit, expressed as

\begin{equation}
    \label{eq1}
    \text{HJD(eclipse)} = \text{HJDo} + \mathrm{P} \times \mathrm{E},    
\end{equation}to determine the values and associated errors for the epoch HJDo, the orbital period P, and the standard deviation in the observed-calculated ($\overline{O-C}$) value. In certain instances, epochs related to the secondary minimum were also incorporated into the fitting process, resulting in a half-integer cycle count.

This procedure was applied in all cases, including those that already had published ephemeris data in the VSX catalogue, as these often have low precision that could be improved or modified by our analysis. In these cases, we normally included the epoch HJDo mentioned in the VSX catalogue as one additional data point in our fit (Eq.\,\ref{eq1}), assigning them E = 0 if their epochs were earlier than our first eclipse recorded from \textit{Gaia} Alerts.

\section{Results and discussion}


The DNe for which we obtained an orbital period can be subdivided into two groups: The first group, listed in Table\,\ref{Tab.11}, consists of 27 DNe with raw ephemeris data provided in the VSX catalogue, which we improved or modified. In this table, we present the coordinates, orbital periods, and subtypes\footnote{The complete list of variable star type designations in VSX can be found at \href{https://www.aavso.org/vsx/index.php?view=about.vartypes}{https://www.aavso.org/vsx/index.php?view=about.vartypes}.} from the VSX catalogue, along with the results of our analysis of \textit{Gaia} data in quiescence. Specifically, we include the number of eclipse observations (N) used in our fit (Eq.\,\ref{eq1}), the approximate total time interval covered ($\Delta$T), the first epoch (HJDo), and the orbital period (P) with their respective errors, as well as the standard deviations ($\overline{O-C}$).  

For most cases, the newly determined periods are highly consistent with the periods listed in the VSX catalogue, which demonstrates the validity of our methods. The only three exceptions are Gaia22dap[ID 83], Gaia19efo[ID 71], and Gaia23bdv[ID 56].
For the first two, the period published in the VSX catalogue is nearly exactly half of that obtained with our methods. Both correspond to systems with clear ellipsoidal variations, and therefore it is likely that the two brightness peaks have been mistaken for a single one in the period estimation listed in the VSX catalogue. For Gaia23bdv[ID 56], the published period is nearly three times ours. Although the system also exhibits clear signs of ellipsoidal variation, the reason for this discrepancy is not as clear to us.

A second group of a total of 75 targets comprises DNe with previously unknown orbital periods. In Table\,\ref{Tab.2}, we provide their identification, coordinates, and subtype according to the VSX catalogue, along with additional data from our research, following the same format as in Table\,\ref{Tab.11}.  Both tables are organised based on the Right Ascension (RA) of the objects.

The cycle numbers E and the epochs HJD of all individual observations of eclipses or brightness minima N used for fitting Eq.\,\ref{eq1} are listed in Table \ref{Tab.A1} for each DN whose ephemerides are given in Tables \ref{Tab.11} and \ref{Tab.2}. Table \ref{Tab.A1} is provided to be included in the analysis of forthcoming eclipse observations or supplementary historical data. The data in Table \ref{Tab.A1} are important because the accuracy of any period determination primarily depends on the total time $\Delta$T covered, even if some of the eclipses do not coincide exactly with the epoch of the true eclipse minimum, as is possible in some of our \textit{Gaia} data. This table is available at the CDS.

\begin{table}[ht]
\centering 
\caption{\label{Tab.A1}Individual \textit{Gaia} Alerts eclipse (or brightness minima) epochs, including the cycle number (E) and the corresponding HJD. The full version of this table is available at the CDS.} 
\begin{tabular}{cccccc}

    \hline
    \hline
    \\[-1.5ex]

    \multicolumn{2}{c}{Gaia23age = ASASSN-13cx}\\
    \hline
    
    \\[-1.5ex]
    E  & HJD - 2\,450\,000 (d)\\  
    \\[-2ex]
    \hline
    \\[-2ex]
    0  & 6901.6982 \\
    910  & 7627.1515 \\
    11604  & 7825.9581 \\
    17882  & 8326.0018 \\
    18416  & 8368.5356 \\
    \\[-1.8ex]
    \hline
    \hline
    \\[-1.5ex]

    \multicolumn{2}{c}{Gaia20bij }\\
    \hline
    \\[-1.5ex]
    E     &       HJD - 2\,450\,000 (d)\\
    \\[-2ex]
    \hline
    \\[-1.5ex]
    0    &   6975.8751\\ 
    565 & 7132.6155 \\
    3295 & 7890.0241 \\
    4055 & 8100.8758 \\
    4743 & 8291.7554 \\
    5231 & 8427.1497\\
    5252 & 8432.9770\\
    5487 & 8498.1632\\
    7422 & 9035.0267\\
    8044 & 9207.5933\\
    8401 & 9306.6237\\
    8653 & 9376.5494\\
    \\[-1.8ex]
    \hline
    \hline
    \\[-1.5ex]

    \multicolumn{2}{c}{Gaia19bfh }\\
    \hline
    \\[-1.5ex]
    E      &      HJD - 2\,450\,000 (d)\\ 
    \\[-2ex]
    \hline
    \\[-2ex]
    0 & 7297.3047\\ 
    1240 & 7606.5748\\ 
    4282 & 8365.2872\\ 
    5407 & 8645.8751\\ 
    5515 & 8672.8018\\ 
    6251 & 8856.3785\\ 
    7473 & 9161.1490\\ 
    7474 & 9161.3990\\ 
    7475 & 9161.6494\\ 
    7476 & 9161.9000\\ 
    7477 & 9162.1500\\ 
    \hline

\end{tabular}
\end{table}


\subsection{Phase-folded light curves}

This study offers a unique opportunity to present a comprehensive collection of phased-folded light curves of typical DNe in quiescence derived from a common high-precision photometric dataset. In addition to the objects listed in Tables\,\ref{Tab.11} and \ref{Tab.2}, we include the light curves for 15 additional DNe from the VSX catalogue with known precise orbital ephemerides. For those, we did not attempt to apply our period search methods. Instead, we simply calculated their light curves from their \textit{Gaia} Alert data using their known ephemerides. 
Eclipses were missing in some cases because \textit{Gaia}, by chance, never observed the corresponding phases. Nevertheless, we included their light curves to demonstrate that \textit{Gaia} data, although sparse, can be effectively combined with typical orbital light curves of DNe in quiescence. Despite significant dispersion in the data, it is possible to determine their orbital periods with high precision, as shown for Gaia17bcx[ID 11] and Gaia19bid[ID 14]. Furthermore, this approach increases the number of systems below the period gap in our light curve collection, which is crucial for the subsequent discussion.

Figure \ref{fig1} shows a total of 117 light curves organised in ascending order according to their orbital periods, ranging from 0.05 to 2.0 days. The different symbols correspond to the three groups mentioned above: DNe with previously known, precise ephemerides from the VSX catalogue (crosses), systems for which the ephemerides were improved by our methods (from Table\,\ref{Tab.11}; open circles), and DNe with new orbital periods (from Table\,\ref{Tab.2}; filled dots). Each light curve in Fig.\,3 includes its identification number (ID) at the beginning of the header, making it easy to locate the corresponding system in Tables \ref{Tab.11} or \ref{Tab.2}.

\begin{figure}[h!]
\centerline{\includegraphics[width=0.9\columnwidth ]{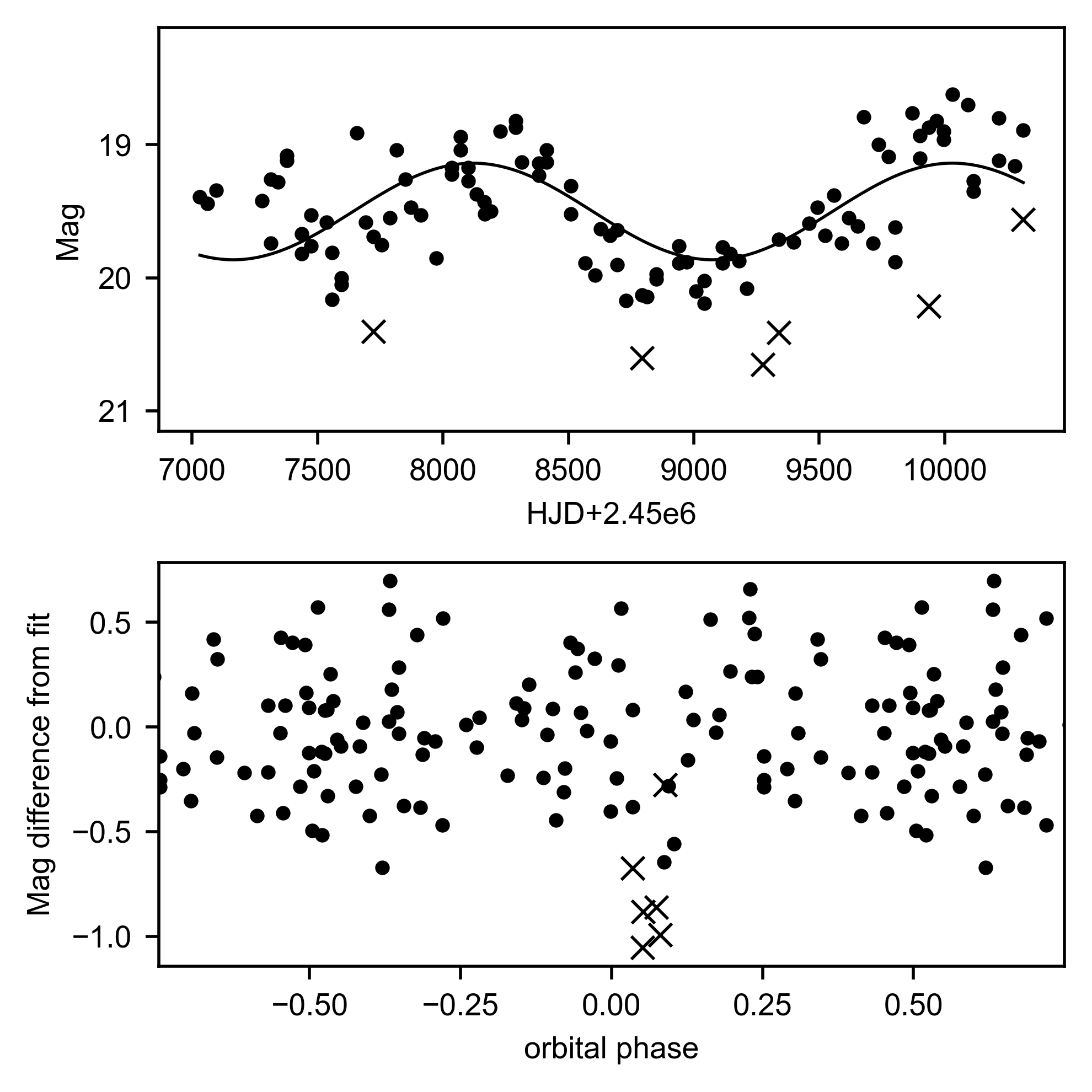}}
    \caption{\label{fig_LT} Light curve of Gaia19auk[ID 54], displaying long-term variations (upper), and phase-folded light curve accounting for these variations (lower). Data points for isolated eclipses  are marked with crosses.}
\end{figure}

Most light curves exhibit a scatter in the range between 0.3 and 1 mag outside the eclipses, which cannot be attributed to observational errors because the photometric accuracy of \textit{Gaia} is better than $\sim$0.03 mag for all targets brighter than 20 mag. The observed scatter can be attributed to the intrinsic variability of CVs known as flickering \citep{Brunch}, or potential long-term variability of DNe in quiescence (\citealt{vogt-vega}; Grawe et al. in prep.). As an example, we show the light curve of the quiescent behaviour of Gaia19auk[ID 54] in Fig. \ref{fig_LT}. There are striking sinusoidal long-term variations in the range 18.5-21 mag, but we can distinguish six isolated data points ---marked with crosses in Fig. \ref{fig_LT}---  where the DN is between 0.5 and 1.0 mag fainter than the surrounding brightness level. Only these observations have been used for the period determination of Gaia19auk[ID 54].

While eclipses are clearly visible in Fig.\,3 for several DNe around phase 0, the precise shape and amplitude of these eclipses remain uncertain in several cases, especially for systems with a short eclipse duration and large amplitude, where the real eclipse minimum may have been missed in the \textit{Gaia} data. We also note a few eclipsing systems (Gaia19cwm[ID 2], Gaia17ber[ID 12], Gaia18dgt[ID 20], and Gaia20dxs[ID 53]) that show a single data point with a low brightness (consistent with an eclipse), which deviates from the expected eclipse time in the phase-folded light curve for the derived period. Whether or not these outlier data points are real remains to be confirmed; there could be error in the calculated period, or false data points could be attributed to other observational effects.

The overall shape of the light curve is significantly influenced by the orbital period. DNe with periods of below $\sim$2\,h ($\sim$0.08\,d) typically display a single peak around phase -0.15, immediately prior to the eclipse. This peak is typically attributed to the movement of the hot spot, which is located where the gas stream from the Lagrangian point L1 collides with the outer edge of the accretion disc, towards the observer. In short-period CVs, the majority of the observed visual flux originates from the accretion disc and the hot spot, as the M-type secondary star is relatively faint, facilitating the detection of this `orbital hump'. The amplitude of this hump ranges between 0 and $\sim$0.4\,mag in the $g$-band, and is probably dependent on the orbital inclination and the mass-transfer rate. 

For cases with longer periods, beginning with Gaia21cuu[ID 26] (P$\sim$2.1\,h), another phenomenon can be observed for some DNe: a double-hump pattern with nearly identical peak brightnesses in the light curve, indicative of ellipsoidal variations produced by the deformed Roche-lobe-filling secondary star. In such cases, the observer can discern two passages per orbit of the apparently largest portion of the donor's stellar surface, implying that the light curve is dominated by the radiation of the  secondary star. This behaviour appears suddenly near the lower edge of the well-known period gap of CVs between $\sim$2 and 3 hours \citep{gansicke, zorotovic,Schreiber24}. In that period range, we can see a mix of light curves. In addition to Gaia21cuu[ID 26], we see this double-hump characteristic in Gaia18dqz[ID 27] (P$\sim$2.2\,h) and Gaia21bbg[ID 30] (P$\sim$2.4\,h), while Gaia20com[ID 28] (P$\sim$2.2\,h) and Gaia17aru[ID 29] (P$\sim$2.3\,h)  still show the unique orbital hump typical of the short-period cases below the period gap. 
Relatively weak orbital humps may be present in Gaia21ddn[ID 31] = V1239 Her (P$\sim$2.4 h) and Gaia20edu[ID 33] (P$\sim$3.2 h), indicating that the donor star, while not the dominant source, may still be causing a weak ellipsoidal modulation in the light curves. This could dilute the effect of the hot spot passage in the light curve, preventing us from detecting a clear orbital hump. In Gaia18dqe[ID 32] (P$\sim$2.8\,h), for example, we cannot see clear evidence of either the orbital hump or ellipsoidal variations.

There are also cases where the light curve remains flat outside the eclipse, and the currently available data do not allow for a definitive classification of these light curves. The DNe exhibiting a flat light curve outside the eclipse with the largest period is Gaia20dau[ID 66] (P$\sim$5.8\,h). However, the absence of the double hump is highly unusual for a system with such a long period, and all DNe with periods longer than that of Gaia20dau[ID 66] exhibit this feature without exception.

Most of our double-hump light curves are symmetric, in the sense that the peak
magnitudes of the two humps are identical within the expected errors. However, there
are a few cases in which the first hump maximum at phase $-0.25$ is significantly
brighter than the second one (in Gaia23bvc[ID 63] and Gaia20cnu[ID 68]), while in other cases the hump maximum at phase $+0.25$ seems to be the brighter one (in Gaia19car[ID 49], Gaia20cmj[ID 55], Gaia21cst[ID 60], Gaia19dvg[ID 61], and Gaia18cuv[ID 65]). All these DNe have orbital periods of between $0.17$ and $0.25$\,d. We can compare this to the results of \citet{bruch2024}, who present phased light curves of 48 eclipsing old novae or nova-like stars within the same period range based on TESS data. In this latter sample, some of the double-hump light curves reveal much stronger asymmetries than our DNe, tending to sometimes simulate a single-hump light curve, but always showing a tiny dip at orbital phase $0.5$, the secondary eclipse. The dominating `single maximum' found by  this last author tends to occur around orbital phase $0.4$ for intermediate polars and around phase $0.7$ for SW Sex stars, raising the question of whether this asymmetry depends on the magnetic field strength. In any case, extreme asymmetries such as those in the sample of  \citet{bruch2024} are not present in our data. We believe that most of our double-hump light curves give the correct period. More detailed future observations will confirm or modify our results.

It seems surprising that DNe with relatively short orbital periods, such as those around or below the period gap, reveal secondary stars that are visible in the visual light curves during their quiescent state. 
Theoretically, in these systems, the donor stars should be late M-dwarfs, whose contribution to the visible flux is expected to be negligible. However, ellipsoidal variations have been detected in other DNe close to the lower edge of the period gap, such as QZ\,Ser (P$\sim$2\,h, \citealt{Thorstensen02-2}). In this last case, the authors found that the secondary star was much hotter than expected for an M-dwarf at that period, and consistent with a spectral type K star strongly enriched in helium. Even more surprisingly, double humps consistent with ellipsoidal variations have also been observed during the quiescent state of a very short-period DN, namely EI Psc $=$ 1RXS\,J232953.9+062814 \citep{Thorstensen02-1}. This DN has a period of 64 minutes, even below the `period bouncer' limit, which represents the minimum period of CVs with stellar companions on the main sequence \citep[e.g.][]{paczynski+sienkiewicz81,gansicke,kniggeetal11}. The donor star is also hotter than expected, which is consistent with helium-enriched models. For the two aforementioned DNe, the authors suggest that the secondary stars were already relatively evolved, having undergone significant nuclear evolution before mass transfer began. This could account for the helium enhancement and the higher temperature, enabling the observation of the double hump caused by ellipsoidal variations at such short periods.

\subsection{The orbital period distribution of DNe}

Our results significantly impact the total population of DNe with known orbital periods. According to the VSX catalogue, there are currently a total of 821 DNe with known orbital periods in the range 0.05 < P < 2.0\,d, which corresponds to the total period range covered by our analysis (Tables\,\ref{Tab.11} and \ref{Tab.2}). 
The lower limit of this range was chosen because DNe with extremely short periods, that is, below the orbital period minimum, are expected to contain evolved secondary stars, such as white dwarfs or helium-enriched stars. 
On the other hand, the upper limit was set by our longest-period object, Gaia21buy[ID 116] (P = 1.98582 days). Interestingly, the VSX catalogue lists only four DNe with longer periods, ranging from 2.5 to 19\,d. We note that those long orbital periods most likely indicate that the donors in those DNe have evolved beyond the main sequence \citep[see e.g.][where several CVs with periods longer than 1 day are assumed to have a subgiant donor]{OgleAtlas2015}. 

\begin{table*}[ht]
\caption{Impact of our work on the number of DNe with known orbital periods.}
\label{Tab.3}      
\centering          
\begin{tabular}
{ccc|cc|cc|cc|c}
\hline 
\\[-1.6ex]
Group & Period range (d) & VSX & \multicolumn{2}{c|}{Tab. \ref{Tab.11}} &  \multicolumn{2}{c|}{Tab. \ref{Tab.2}} & \multicolumn{2}{c|}{Total} & Period range selection \\
\\[-2.5ex]
 &  & N & N & \%  & N & \%  & N & \% &
 \\
 \\[-1.6ex]
 \hline 
 \\[-1.2ex] 
A & 0.05 - 0.1 & 536 & 8 & 1.5 & 11 & 2.1 & 19 & 3.5 & below the period gap\\[0.5ex] 
B & 0.1 - 0.13 & 15 & 0 & 0.0 & 1 & 6.7 & 1 & 6.7 & 
within the gap\\[0.5ex]
C & 0.13 - 0.42 & 227 & 16 & 7.0 & 51 & 22.5 & 67 & 29.5 & above the gap, but <10\,h\\[0.5ex]
D & 0.42 - 2.00 & 16 & 3 & 18.8 & 12 & 75.0 & 15 & 93.8 & > 10\,h\\[0.5ex]
Total & 0.05 - 2.00 & 794 & 27 & 3.4 & 75 & 9.4 & 102 & 12.8 & \\[0.5ex]
\hline
\end{tabular}
\tablefoot{This table presents the increase in the total number (N) of DNe with known orbital periods as a result of the present study, along with approximate percentages reflecting our substantial contribution.}

\end{table*}


Out of the 821 DNe with known periods in this range listed in the VSX catalogue, we have improved the orbital period in 27 cases (Table \ref{Tab.11}). This leaves a total of 794 DNe with previously known orbital periods that were not considered in our analysis.
Table\,\ref{Tab.3} displays the total number of DNe with periods listed in the VSX catalogue and those identified in our study, along with the percentage contribution of our new measurements. In addition to the overall sample, we provide these numbers and percentages broken down into four period ranges of interest: (A) below the period gap, (B) within the period gap, (C) above the gap with periods below 10\,h, and (D) with periods larger than 10\,h. The boundaries for the period gap were taken from the most recent estimation in the literature \citep{Schreiber24}, that is, between 147 and 191 min. Our contribution to targets with short periods (below and within the period gap) is relatively modest (3.5$\%$ and 6.7$\%$, respectively). However, in the first group above the period gap (with periods below 10\,h), we have added $\sim$22.5$\%$ to the previously known population, while we have almost doubled the number of previously known periods above $\sim$10\,h, with an increase of $\sim$93.8$\%$. In total, the number of all DNe with known orbital periods, not limited to eclipsing ones but encompassing the entire population, now amounts to 896, to which the 75 new orbital periods presented here (Table \ref{Tab.2}) contribute with 9.4$\%$ to the total population of DNe with known periods. 

The relatively large contribution to newly discovered DNe with long periods as opposed to those with short periods could be partly attributed to the nature of ground-based photometry. Short-period CVs often exhibit two or more eclipses within a single observing night, facilitating the prompt determination of their orbital period. In contrast, for longer-period systems (especially those with periods close to 24 hours), ground-based observations typically only capture segments of the phased light curve over successive nights. This results in large gaps between observations, requiring extensive additional observations to accurately determine the orbital period. 
Furthermore, the majority of DNe below and within the period gap belong to the SU UMa and WZ Sge subtypes. All of these systems exhibit superoutbursts characterised by superhumps, semi-periodic variations during their brightest outburst stages (see \citealt{vogt74} and \citealt{vogt2023}), which are easily observable even by advanced amateur astronomers. On the other hand, \cite{stolz} showed that the superhump period of a DN is correlated with its orbital period \citep[for a modern presentation of this correlation, see][]{otulakowska}. We suggest that many orbital periods of the DNe listed in the VSX catalogue were inferred from this relationship rather than being directly observed, which may account for the relatively large number of DNe with short periods in this catalogue. As mentioned above, superhumps are present in all SU UMa and WZ Sge stars regardless of orbital inclination. While they offer valuable estimations of the orbital period, these estimations must be validated through photometric or spectroscopic observations in the quiescent state.


\begin{figure}[h!]
\centerline{\includegraphics[width=1\columnwidth ]{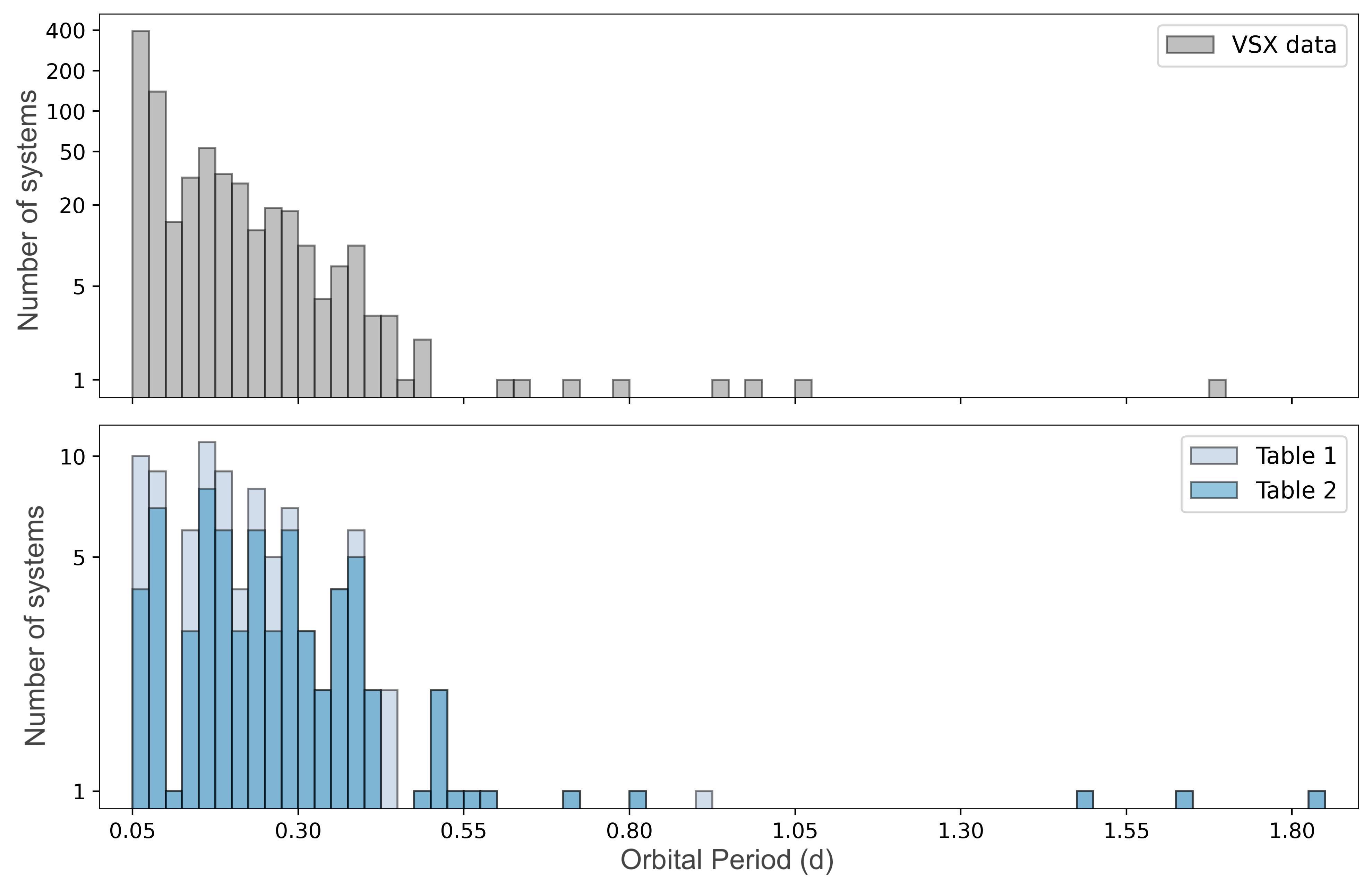}}
    \caption{\label{fig.histo}Period distribution of the known periods from 794 objects retrieved from the VSX catalogue (top) alongside the period distribution of the discovered periods from our analysis (bottom) as presented in Tables\,\ref{Tab.11} and \ref{Tab.2} (light and dark blue, respectively). The vertical axis is on a logarithmic scale.}
\end{figure}

In Fig.\ref{fig.histo} we compare the distribution of orbital periods between the 794 DNe from the VSX catalogue with periods between 0.05 and 2 days (upper panel) with those studied here (lower panel, highlighting systems from Tables \ref{Tab.11} and \ref{Tab.2}). For both populations, a decrease in the number of systems around the location of the period gap (P$\sim$0.10\,d) can be observed. This decrease is more pronounced in the systems studied here, although we note that the exact location of the period gap boundaries is still not well determined \citep[e.g.][]{Schreiber24}.
There is also a general trend showing a decrease in the number of systems as we move from the shortest periods towards systems with  longer periods, up to around 0.55 days, with relatively very few systems above that period. The contribution of our new sample to the total population is particularly evident for DNe with periods longer than 0.4 days.

\section{Conclusions}

Our study is designed to answer the crucial question of whether accurate orbital periods can be derived on timescales of hours from a database with photometric data of high quality, but characterised by irregular time intervals, including long gaps between weeks and several years. Here, we demonstrate that this is feasible using the \textit{Gaia} space telescope database of successful determinations of precise eclipse ephemerides for mostly faint, previously unstudied DNe in quiescence, down to a magnitude of $g\sim\,$20 mag, despite the sparse time distribution of these data.

We present two methods for selecting DN targets with high orbital inclination, deriving new orbital periods in 75 cases and improving or modifying previous ephemerides in another 27 cases, all falling within the range of 0.05\,d < P < 2.0\,d. The similarity between the periods obtained for 24 of these 27 DNe with available ephemerides demonstrates the validity of our methods.
Our results significantly enrich the dataset in the VSX catalogue, and constitute an increase of 9.4$\%$ in the total population of DNe with published periods, impacting the entire DNe population, and not limited to eclipsing DNe. Our contribution is particularly notable for DNe with long orbital periods (P > 0.44\,d), where we almost double the number of previously known periods.

We also present phase-folded light curves for a total of 117 DNe using \textit{Gaia} data, sorted by orbital period. For most systems with periods shorter than 2 hours, we observed the characteristic orbital hump immediately preceding the eclipse caused by the passage of the hot spot. For systems with periods of longer than 3 hours on the other hand, we note the presence of ellipsoidal variations of the distorted secondary star, characterised by two maxima per orbital cycle. The transition between these distinct light-curve patterns occurs around the period gap ($\sim$2-3\,h). However, this transition is not abrupt, and a mixture of systems can be found around these orbital periods. We speculate that the prevalence of DNe with ellipsoidal variations at short periods might be caused by donors that were significantly evolved before the CV phase, as has been proposed to explain these variations in QZ\,Ser. These systems therefore hold promise for future investigations.

Finally, we have established a solid foundation for determining orbital periods using upcoming databases, such as that of the \textit{Vera Rubin} Observatory (Large Synoptic Survey Telescope, LSST). According to \cite{verarubin}, this telescope will scan the entire accessible sky approximately every 4 days over a period of 10 years, providing new opportunities to advance DN research. Future eclipse observations will be combined with the epochs in Table \ref{Tab.A1}, enabling precise orbital ephemerides in the near future and providing a foundation for further studies.

\section{Data availability}

Table \ref{Tab.A1} is only available in electronic form at the CDS via anonymous ftp to \href{http://130.79.128.5/}{cdsarc.u-strasbg.fr (130.79.128.5)}  or via \href{http://cdsweb.u-strasbg.fr/cgi-bin/qcat?J/A+A/.**}{http://cdsweb.u-strasbg.fr/cgi-bin/qcat?J/A+A/.}\\

\begin{acknowledgements}
     We would like to thank the team of the Variable Star Index (VSX) of AAVSO, for the very efficient and permanently actualised information on variable stars, also offering direct access to the \textit{Gaia} Alerts data. NV and MZ acknowledge financial support by Centro de Astrofísica de Valparaíso (CAV).
     A.R.-J. acknowledge funding from ANID-Subdirección de Capital Humano/Doctorado Nacional/2022-21221841. We also would like to thank the referee Albert Bruch for valuable suggestions which improved considerably the earlier version of this article.
\end{acknowledgements}

\bibliographystyle{aa}
\bibliography{ref}

\begin{appendix} 


\section{Additional tables}

Here we present the refined ephemeris for the 27 DNe with estimated periods from the VSX catalogue (Table\,\ref{Tab.11}) and for the 75 Dne for which the period was first determined in this study (Table\,\ref{Tab.2}).


\setlength{\tabcolsep}{4pt}
\begin{table}[!h]
    \caption{DNe with previously known orbital periods.}
    \label{Tab.11}
    \centering
    \tiny
    \begin{tabular}{ccccccc|ccccc}
    \hline
    
    \\[-1.4ex]
\multicolumn{7}{c}{\small{VSX data}} & \multicolumn{5}{c}{\small{New ephemeris}} \\
    \\[-1.4ex]
    \hline
    \\[-2.0ex]
    \textit{Gaia} Alerts & Star name & ID &\multicolumn{2}{c}{Coord.\,(2000)} &  P(VSX)\,(d) & Subtype & N & $\Delta$T\,(yrs) & HJDo-2\,450\,000(d) & P\,(d) & $\overline{\mbox{O-C}}$\,(d)\\
    \\[-2.0ex]
& & & $\alpha$(h) & $\delta$($^{o}$) & & & & & & & \\
    \\[-1.5ex]
    \hline
    \\[-2.5ex]
23age & ASASSN-13cx & 21 & 00 02 22.4 & +42 42 13 & 0.079650 & UGSU+E & 5 & 4.0 & 6901.6980(3)  & 0.07965015(2) & 0.0003 \\
20ecl & ASASSN-18aan & 38 & 00 46 08.0& +62 10 05 & 0.14941 & UGSU+E & 4 & 1.80 & 8059.1976(35)  & 0.149445(1) & 0.0044\\
21fbh & ASASSN-18yi  & 46 & 01 21 52.2 & -33 56 16  & 0.16825 & UF+E & 8 & 6.5 & 7040.5018(41) & 0.1682264(5) &  0.0064\\
22dap &  & 83 & 02 25 41.3 & +53 26 55 & 0.148897 & UG+E & 8 & 7.3 & 7308.0714(98) & 0.2978016(15) & 0.0111 \\
19bcd & KW And & 76 & 02 35 18.0 & +41 14 02 & 0.27297 & UGSS+E & 8 & 5.0 & 7026.644(12)  & 0.273196(3)  & 0.0165\\
22eit &  & 67  &  02 40 40.8 & +54 59 58 & 0.24025 & UG+E & 12 & 7.8 & 7309.254(3)  & 0.2402498(5)  & 0.0065\\
20dsw & AY For & 18  &  02 42 34.8 & -28 02 44 & 0.07460 & UGSU+E & 12 & 17.7 & 2577.4367(3)  & 0.074614832(6) & 0.0007\\
20ein & LT Eri &  47 & 04 07 14.8 & -06 44 25 & 0.17017 & UG+E & 12 & 16.2 & 2949.1477(15)  & 0.17020386(10) & 0.0029\\
17anx & ASASSN-14ka &  50 &  04 20 39.8 & -62 45 01  & 0.17716 & UG+E & 6 & 2.0 & 6886.9834(14)  & 0.1771601(5) & 0.0023\\
19auk & ASASSN-15pw &  54  &  04 46 39.4  & -51 32 54 & 0.1834 & UG+E & 7 & 8.2 & 7316.5894(27) & 0.18338117(26) & 0.0038\\
21cmv &  & 6  &  05 22 09.7 & -35 05 30 & 0.0622 & UG+E & 7 & 10.9 & 5913.4370(5)  &  0.062193493(11) & 0.0006\\
18atj &  &  103 & 06 27 24.4 & +16 13 30 & 0.434537 & UG+E & 12 & 5.4 & 7299.621(14)  &  0.434514(5) & 0.0205\\
20air & NSV 4618 &  9 &  09 45 51.0 & -19 44 01 & 0.065769 & UGSU+E & 8 & 11.8 & 4884.1338(16)  & 0.06576932(3) & 0.0017\\
20ezk & ASASSN-15aa &  94 & 10 49 26.0 & -21 47 36 & 0.375540 & UGSS+EW & 12 & 6.1 & 7109.980(7)  & 0.3755385(16) & 0.0095\\
22ajs & ZTF18aabefy1 & 10  & 11 52 07.0 & +40 49 48 & 0.067721 & UGSU+E & 12 & 11.7 & 4879.5386(2) & 0.067749690(6) & 0.0005\\
21czo & & 7  & 13 25 36.1 & +21 00 37 & 0.062238 & UG+E & 3 & 11.5 & 5252.8343(25) & 0.06238490(6) & 0.0027\\
19cwm & ASASSN-19qy  &  2 & 16 27 16.8 & +04 06 03  & 0.05995 & UGWZ+E & 6 & 2.4 & 8900.7886(7) & 0.05989249(8) & 0.0011\\
20fjv & NSV11599 &  102  & 18 56 51.5 & +53 40 09 & 0.4294162 & UG+E & 10 & 5.1 & 7343.2488(84)  & 0.429420(3) & 0.0099\\
22beu & ZTF19aaylohn  &  74 & 19 06 28.2 & +08 36 50 & 0.2622809 & UGSS & 13 & 6.7 & 7455.232(3)  & 0.2622820(7) & 0.0074\\
19asr & KIC 8625249 & 43  & 19 27 48.5 & +44 47 25 & 0.1653077 & UG+E & 5 & 5.7 & 6206.0841(37)  & 0.1653123(4) & 0.0040\\
19efo & ASASSN-19xi &  71  & 19 28 30.0 & -74 10 53 & 0.1248492 & UG+E & 12 & 4.7 & 7112.923(4) & 0.2496973(10) & 0.0072\\
20fii &  & 112  & 19 37 24.9 & +15 25 57 & 0.9186710 & UG+E & 7 & 4.5 & 7076.578(23)  &  0.918659(20) & 0.0390\\
20com & ASASSN-15pb   & 28 & 20 14 23.0 & -63 37 58 & 0.09329 & UG+E & 9 & 6.3 & 7312.3498(22) & 0.09329004(11) & 0.0032\\       
23bdv & ZTF19acxzdwq   & 56 & 20 42 56.3 & +12 44 58 & 0.59747 & UG & 15 & 7.5 &  7370.128(3) &  0.1937450(4) & 0.0082\\
17bms &    & 35 & 21 28 20.1 & +55 28 48 & 0.1446067 & UG+E & 7 & 4.6 & 7306.0431(18)  & 0.1446087(3) & 0.0032\\
22dfe & ASASSN-13cl   & 59  & 21 38 05.0 & +26 38 20 & 0.20219 & UG+E & 16 & 10.1 & 6551.8092(29) & 0.20216428(24) & 0.0053\\  
20eja & MGAB-V229   & 36 & 21 55 01.4 & +49 58 09 & 0.146830 & UG+E & 3 & 1.9 & 7023.1635(26) &  0.1468361(9) & 0.0029\\

\hline
\end{tabular}
\onecolumn
\tablefoot{The first seven columns refer to data from the VSX catalogue along with the assigned ID number, which was introduced to facilitate the search for each system in Figure \ref{fig1}, where they are ordered by orbital period. N corresponds to the total number of individual eclipses or brightness minima included in the ephemeris calculation; $\Delta$T indicates the approximate total time interval covered. HJDo and P denote the ephemeris, and $\overline{\mbox{O-C}}$ represents the mean individual eclipse time deviations from the linear fit.}

\end{table}



\onecolumn
\setlength{\tabcolsep}{4pt}
\tiny
\begin{longtable}{cccccccccccc}
\caption{\label{Tab.2} DNe with new orbital periods.}\\

\hline

\\[-1.4ex]

\multicolumn{6}{c}{\small VSX data} & \multicolumn{5}{c}{\small New ephemeris} \\ \cline{8-12}
  \cline{1-6}
  \\[-1.4ex]
\textit{Gaia} Alerts & Star name & ID & \multicolumn{2}{c}{Coord.\,(2000)} & Subtype & & N & $\Delta$T (yrs) & HJDo\,-\,2\,450\,000\,(d) & P\,(d) & $\overline{\mbox{O-C}}$\,(d)\\
 &  &  & $\alpha$\,(h) & $\delta$\,($^{o}$) & & \\

 \\[-1.9ex]
\hline
\\[-1.9ex]
\endfirsthead 

\caption{continued}\\
\hline
\\[-1.4ex]
\multicolumn{6}{c}{\small VSX data} & \multicolumn{5}{c}{\small New ephemeris} \\ \cline{8-12}
  \cline{1-6}
  \\[-1.4ex]
\textit{Gaia} Alerts & Star name & ID & \multicolumn{2}{c}{Coord.\,(2000)} & Subtype & & N & $\Delta$T (yrs) & HJDo\,-\,2\,450\,000\,(d) & P\,(d) & $\overline{\mbox{O-C}}$\,(d)\\
 &  & & $\alpha$\,(h) & $\delta$\,($^{o}$) & & \\
 \\[-1.9ex]
\hline
\\[-1.9ex]
\endhead 
\hline
\endfoot 

20bij & & 79 & 00 26 09.8& +54 32 42 & UG: & & 12 & 4.6 & 6975.865(4)  & 0.2774391(7) & 0.0068\\
19bfh &  & 70 & 00 37 17.9 & +52 25 12 & UG & & 11 & 5.1 & 7297.307(3) & 0.2494106(4) & 0.0038 \\

17bjn &  & 73& 00 53 47.3 & +40 55 49 & UG & & 6 & 5.4 & 7049.058(5) & 0.2616035(11) & 0.0069 \\

20ane & ASASSN-15sy & 42 & 01 11 13.5 & -44 41 34 & UG & & 7 & 4.4 & 8551.207(3) & 0.1617499(4) & 0.0045 \\

17aru &  & 29 & 02 04 01.1 & +43 41 32 & UG & & 6 & 6.0 & 7452.6638(4) & 0.09620147(5) & 0.0007 \\

18cnp & ASASSN-14gl & 45 & 02 05 00.4 & +45 05 38 & UGZ & & 6 & 4.2 & 7023.142(3) & 0.1673839(4) & 0.0038 \\

19bbb & AT 2019cbz & 44 & 03 28 15.9 & +61 25 17 & UG & & 9 & 1.0 & 7673.777(4) & 0.166856(3) & 0.0079 \\

19ery & & 106 & 04 09 59.7 & +69 14 39 & UG: & & 11 & 3.8 & 7979.913(7) & 0.515631(5) & 0.0149 \\

18dqe & & 32 & 04 58 32.1 & +32 49 01 & UG & & 7 & 9.0 & 7029.0053(13) & 0.11861707(9) & 0.0023 \\

21apj & ASASSN-18abu & 52 & 05 17 04.5 & -25 31 03 & UG+E & & 7 & 6.6 & 7165.976(8) & 0.1825278(7) & 0.0081 \\

22ddq & & 91 & 06 04 26.8 & -11 57 45 & UG & & 14 & 6.4 & 7330.293(4) & 0.3695405(12) & 0.0089 \\

17ber & & 12 & 06 33 04.8 & +03 24 23 & UG|QSO & & 5 & 5.0 & 7702.5855(5) & 0.06964211(4) & 0.0008 \\

21bge & & 15 & 06 41 27.5 & +13 48 16 & UGZ & & 6 & 2.1 & 8718.4767(5) & 0.07211448(8) & 0.0008 \\

19ajr & & 16 & 07 00 17.6 & -34 20 30 & UG+NL/VY: & & 8 & 7.0 & 7617.9064(5) & 0.073589950(22) & 0.0010 \\

20edh & & 89 & 07 01 01.7 & +12 18 33 & UG & & 16 & 5.9 & 7131.730(3) & 0.3354595(9) & 0.0063 \\

19aoo & ASASSN-16an & 40 & 07 08 16.6 & -12 45 45 & UG & & 15 & 6.5 & 7095.980(2) & 0.1571464(3) & 0.0055 \\

18dmg & & 95 & 07 10 43.6 & +04 19 30 & UG: & & 10 & 6.5 & 7099.250(14) & 0.379773(3) & 0.0167 \\

18dqz & & 27 & 07 16 10.3 & -01 30 50 & UG & & 6 & 4.7 & 7653.664(3) & 0.0913635(3) & 0.0048 \\

21bct & & 19 & 07 21 34.3 & -07 21 14 & UG & & 6 & 6.6 & 7331.8585(11) & 0.07481684(6) & 0.0014 \\

19dvg & & 61 & 07 22 19.0 & +46 49 04 & UG & & 9 & 6.2 & 6987.247(3) & 0.2064332(5) & 0.0047 \\

20eac & & 85 & 07 23 20.0 & -01 37 24 & UG & & 11 & 4.2 & 8205.519(5) & 0.3077270(13) & 0.0079 \\

22cnl & & 41 & 07 34 24.1 & -76 37 58 & UGZ & & 5 & 6.4 & 7748.3429(14) & 0.15946083(13) & 0.0015 \\

21fpm & & 82 & 07 41 08.9 & -10 55 17 & UG|AGN & & 12 & 6.9 & 7333.348(6) & 0.2962373(11) & 0.0117 \\

20beu & & 62 & 07 52 56.9 & -39 05 37 & UG+E & & 9 & 7.5 & 7089.474(3) & 0.2130102(5) & 0.0052\\

20cnu & & 68 & 07 57 47.2 & -15 05 44 & UG & & 11 & 5.4 & 7383.329(6) & 0.2466667(12) & 0.0091\\

23bvc & ZTF18aabnrpx & 63 & 08 04 08.0 & +08 37 15 & UG & & 8 & 6.3 & 7174.275(5) & 0.2251499(7) & 0.0068 \\

21bbg & & 30 & 08 19 28.8 & -57 26 06 & UG+E: & & 11 & 7.5 & 7058.2837(13) & 0.09968375(8) & 0.0024 \\

19car & ASASSN-17as & 49 & 08 20 00.1 & -29 21 45 & UG & & 7 & 3.4 & 7614.912(7) & 0.1740800(18) & 0.0108 \\

18dgt & & 20 &  08 39 59.8 & +74 02 27 & UG+E & & 6 & 3.5 & 7209.3340(8) & 0.07961705(9) & 0.0011 \\

19fkr & &39 & 08 51 18.9 & -42 39 49 & UG & & 6 & 6.3 & 7718.1572(11) & 0.15508925(9) & 0.0012 \\

19dhr &  & 34 & 08 54 39.6 & -32 24 48 & UG & & 12 & 4.8 & 7563.975(3) & 0.1428998(3) & 0.0035 \\

19brc & & 48 & 09 05 57.2 & -38 28 09 & UG & & 13 & 5.0 & 7159.940(3) & 0.1739515(4) & 0.0053 \\

19eql & ASASSN-18ea & 108 & 09 23 11.3 & -16 32 31 & UG & & 9 & 3.5 & 7367.032(9) & 0.563360(9) & 0.0169 \\

21bpp & & 115 & 09 37 07.9 & -60 14 13 & UG & & 11 & 6.6 & 7503.548(33) & 1.82552(5) & 0.0739 \\

20anw & & 23 & 10 17 50.4 & -50 15 40 & UG+E & & 6 & 6.6 & 7129.5098(10) & 0.08336320(6) & 0.0015 \\

20dau & ASASSN-19gm & 66 & 10 24 16.9 & -42 53 53 & UG+E & & 8 & 5.8 & 7344.2801(14) & 0.2399637(3) & 0.0023 \\

20bhb & & 100& 10 57 51.2 & -53 24 25 & UG & & 16 & 5.1 & 7129.426(4) & 0.4046580(14) & 0.0074 \\

19eux & & 77 & 10 59 24.5 & -62 42 27 & UG & & 10 & 5.7 & 7402.587(6) & 0.2754071(12) & 0.0095 \\

18dge & & 80 &  11 23 59.5 & -52 14 18 & UG & & 14 & 5.3 & 7509.801(7) & 0.2913592(15) & 0.0123 \\

20edu & & 33 & 11 41 34.7 & -50 48 29 & UG+E & & 7 & 7.3 & 6973.3129(8) & 0.13442575(6) & 0.0012 \\

19akh & & 98 & 13 01 35.7 & -05 29 38 & UG & & 17 & 9.4 & 7065.004(6) & 0.3903374(12) & 0.0141 \\

19aee & & 88 & 13 11 18.9 & -52 15 14 & UG & & 9 & 5.2 & 7025.768(3) & 0.3346760(8) & 0.0040 \\

18boh & & 109 & 13 55 49.4 & -49 46 09 & UG & & 8 & 5.6 & 7200.673(9) & 0.578325(5) & 0.0127 \\

19fnf & & 69 &  14 18 30.8 & -68 35 44 & UG: & & 13 & 5.1 & 7354.798(4) & 0.2485503(9) & 0.0081 \\

22ace & & 97 & 14 19 54.9 & -51 25 15 & UG: & & 7 & 7.5 & 7046.761(3) & 0.3839405(9) & 0.0053 \\

19aag & & 110 & 15 06 18.7 & -65 08 18 & UG & & 14 & 6.6 & 6982.265(7) & 0.716905(4) & 0.0142 \\

21bwq & ASASSN-21fs & 93 &15 50 22.5 & -62 16 58 & UG & & 15 & 6.4 & 8012.039(4) & 0.3720838(13) & 0.0091 \\

18bgv & & 72 & 16 29 54.7 & -74 41 51 & UG & & 10 & 5.4 & 7521.034(3) & 0.2607071(8) & 0.0067 \\

21cst & & 60 & 17 18 59.8 & -73 58 16 & UG & & 19 & 7.0 & 6978.753(5) & 0.2037822(7) & 0.0120 \\

21apv & & 24 & 17 57 58.9 & +24 24 27 & UGSU:+E & & 6 & 6.5 & 7476.9627(13) & 0.08501096(6) & 0.0015 \\

22dum & ZTF18abstqob & 51 & 18 23 47.0 & -16 30 03 & UGZ & & 6 & 4.9 & 8070.548(3) & 0.1800017(5) & 0.0037 \\

19bwi & & 86 & 18 42 53.4 & -03 27 34 & UG & & 7 & 4.6 & 7651.287(2) & 0.3079671(10) & 0.0044 \\

21anl & ASASSN-15ea & 105 & 18 50 50.6 & +40 44 06 & UG & & 20 & 7.5 & 7376.409(5) & 0.5137853(18) & 0.0120 \\

21buy & ASASSN-21dr & 116& 18 52 49.6 & -24 50 28 & UG & & 15 & 4.4 & 8023.586(24) & 1.98582(6) & 0.0561 \\

17aqz & & 90 & 18 53 33.5 & +22 35 59 & UG: & & 17 & 5.6 & 7384.661(4) & 0.3590382(12) & 0.0082 \\

20dxs & & 53 & 19 00 42.7 & -27 20 55 & UG & & 7 & 5.5 & 8257.394(3) & 0.1828514(4) & 0.0051 \\

20dbv & KIC 4450058 & 101 & 19 07 59.5 & +39 34 14 & UG & & 15 & 6.6 & 7022.865(9) & 0.407165(2) & 0.0181 \\

20dzz & ASASSN-20kh & 75 & 19 17 12.8 & -14 27 17 & UG & & 18 & 3.5 & 8076.634(6) & 0.264211(2) & 0.0114 \\

19fgn & & 104 & 19 22 54.6 & -18 38 36 & UG & & 11 & 6.5 & 7457.97989(9) & 0.485335(3) & 0.0149 \\

20col & & 99 & 19 38 52.3 & +30 54 12 & UG & & 13 & 6.4 & 7504.395(6) & 0.393722(2) & 0.0123 \\

20akg & & 92 & 20 11 01.2 & +28 17 23 & UG & & 12 & 6.4 & 7469.647(7) & 0.3712763(14) & 0.0111 \\

18dvx & & 96 & 20 13 59.3 & +36 54 15 & UG & & 13 & 5.6 & 7339.907(9) & 0.380905(3) & 0.0211 \\

22chc & ZTF18achegif & 111 & 20 19 34.2 & +06 13 55 & UG+E & & 10 & 6.8 & 7374.415(12) & 0.801344(8) & 0.0290 \\

16bxp & & 78 & 20 22 11.6 & +31 49 06 & UG: & & 10 & 3.3 & 8515.347(6) & 0.2766039(23) & 0.0095 \\

18buk & & 84 & 20 25 42.6 & +26 20 44 & UG & & 16 & 4.2 & 7768.952(3) & 0.3051541(12) & 0.0079 \\

16aik & AT 2016dvk & 37 & 20 43 59.6 & +42 03 25 & UG & & 5 & 4.1 & 8128.792(3) & 0.1476076(5) & 0.0045 \\

18dlr & & 57 & 20 52 58.0 & +45 54 19 & UG: & & 7 & 5.2 & 7776.694(4) & 0.1989203(9) & 0.0077 \\

20cmj & & 55 & 20 53 50.2 & +39 02 26 & UG: & & 13 & 7.7 & 7063.650(3) & 0.1892539(3) & 0.0044 \\

22btj & ZTF18acrnput & 114 & 21 03 16.5 & +31 49 14 & UG & & 26 & 7.9 & 7449.719(18) & 1.640711(22) & 0.0552 \\

21cuu & & 26 &  21 15 48.8 & -37 44 43 & UG & & 18 & 7.0 & 7482.7639(7) & 0.08863238(4) & 0.0018 \\

22eer & MGAB-V1027 & 81 & 21 38 06.6 & +32 19 42 & UG+E & & 11 & 6.4 & 7830.4813(22) & 0.2939460(5) & 0.0046 \\

21apu & & 58 & 21 40 58.4 & +28 28 38 & UG & & 5 & 5.5 & 7836.485(4) & 0.1994121(7) & 0.0051 \\

23bbd & ZTF19aassvxn & 107 & 22 00 33.5 & +37 38 49 & UG & & 19 & 9.0 & 6867.064(7) & 0.5375198(17) & 0.0120 \\

21azp & & 113 & 22 12 13.9 & +61 20 02 & UG: & & 7 & 5.8 & 6889.888(33) & 1.48700(4) & 0.0477 \\

18cuv & ASASSN-18wn & 65 & 22 55 51.3 & -80 45 18 & UG & & 20 & 4.9 & 7213.390(2) & 0.2285728(5) & 0.0066 
\end{longtable}

\section{Phase-folded light curves}

\begin{figure}
    \centerline{\includegraphics[width=18.5 cm]{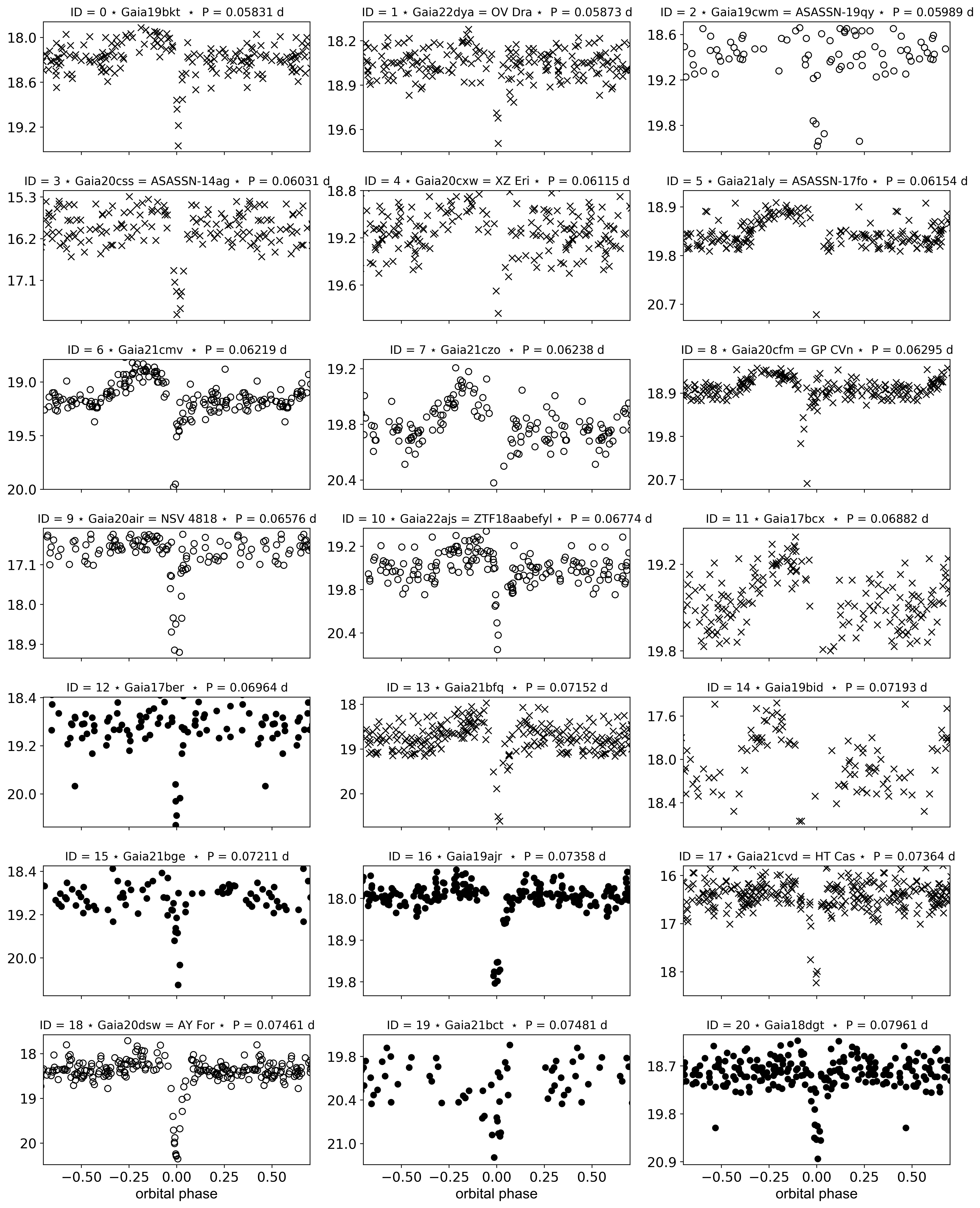}}
    \caption{Phased-folded light curves of DNe, based on Gaia alert data in the quiescent state, sorted by their orbital period in ascending order. The y-axis represents the magnitude. The top of each figure displays their identification number (ID), the GAIA name, and its corresponding period in days. Crosses denote cases with previously known precise ephemeris values listed in the VSX catalogue, not modified by us. Open circles represent objects with periods suggested in the VSX catalogue but improved or modified by our analysis (listed in Table\,\ref{Tab.11}). Those marked with filled dots correspond to systems with new periods determined in this study (Table\,\ref{Tab.2}).}
    \label{fig1}
\end{figure}

In Figure \ref{fig1} we present the phased-folded light curves of DNe, organised by their orbital
period in ascending order.

\setcounter{figure}{0}
\begin{figure}
    \centerline{\includegraphics[width=19 cm]{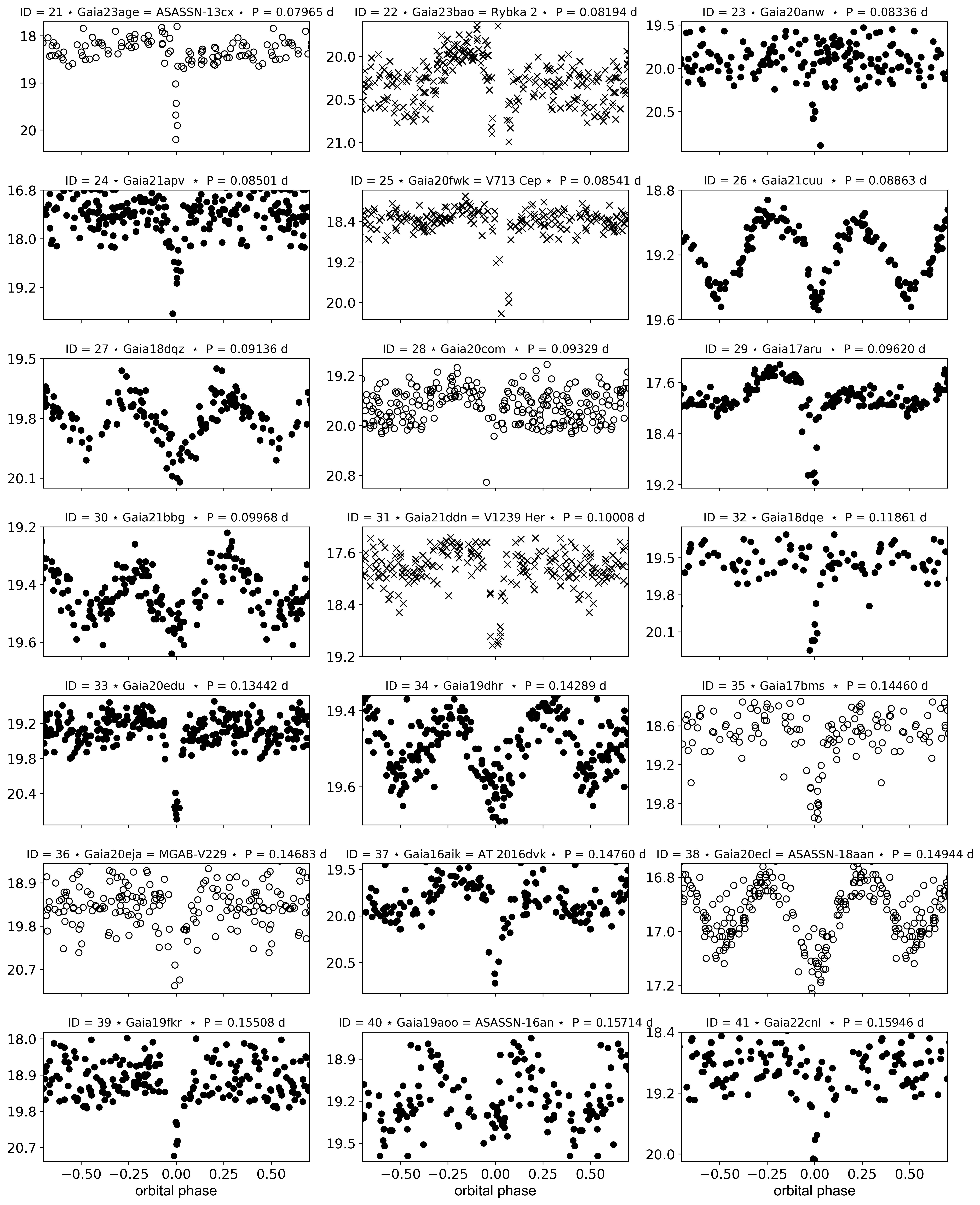}}
    \caption{\textit{– continued}}
    \label{fig2}
\end{figure}

\setcounter{figure}{0}
\begin{figure}
    \centerline{\includegraphics[width=18.5 cm]{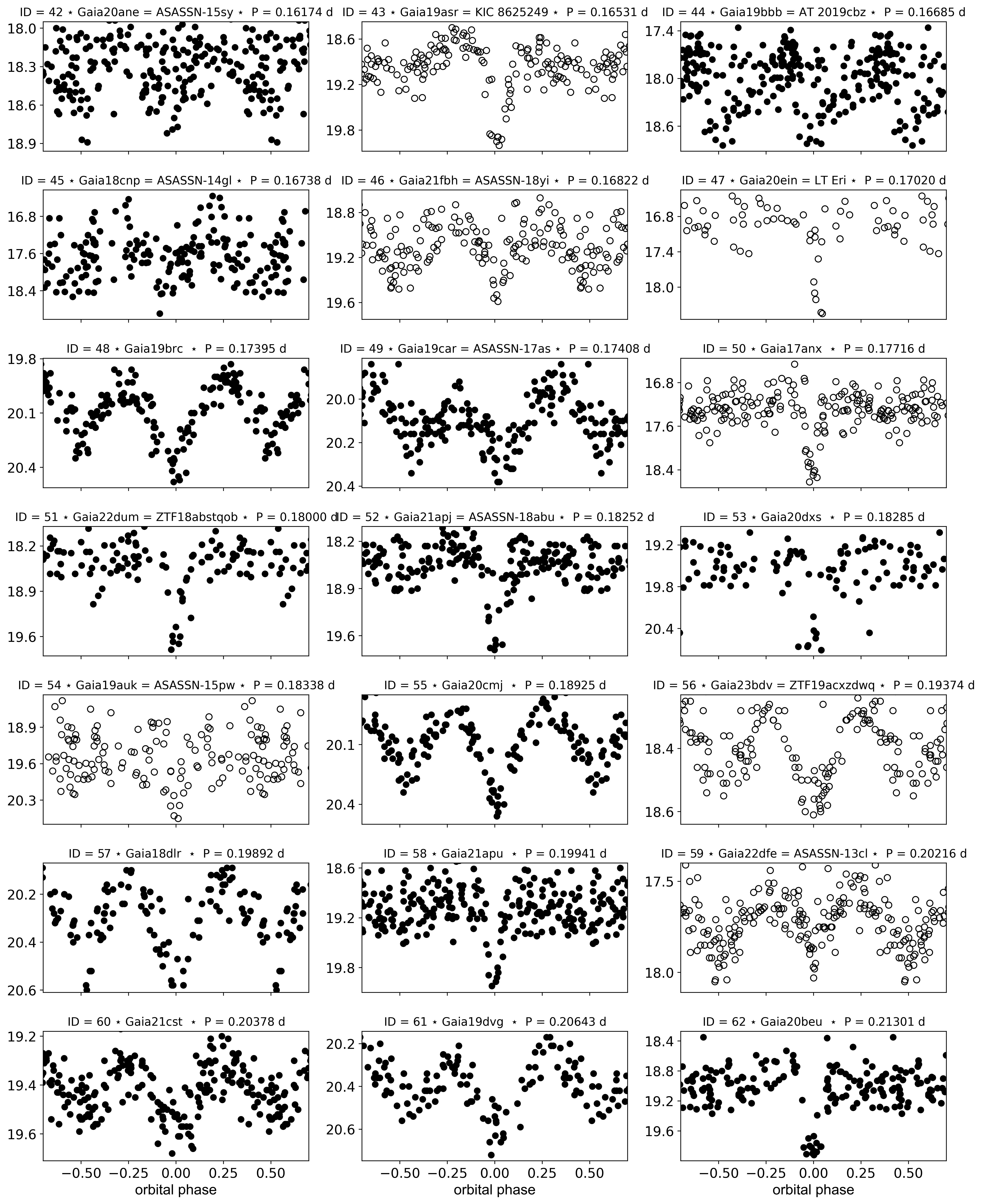}}
    \caption{\textit{– continued}}
    \label{fig3}
\end{figure}

\setcounter{figure}{0}
\begin{figure}
    \centerline{\includegraphics[width=18.5 cm]{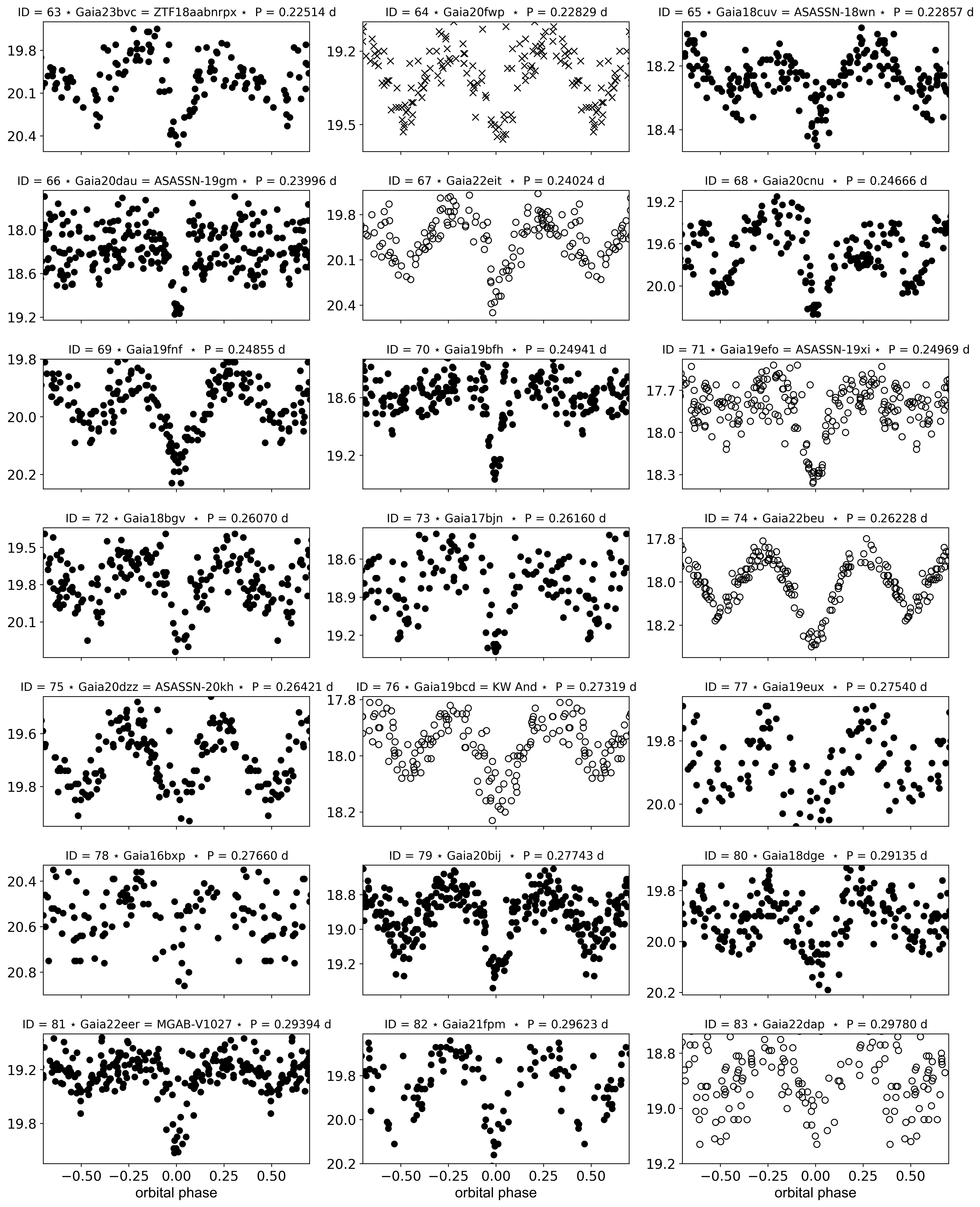}}
    \caption{\textit{– continued}}
    \label{fig4}
\end{figure}

\setcounter{figure}{0}
\begin{figure}
    \centerline{\includegraphics[width=19 cm]{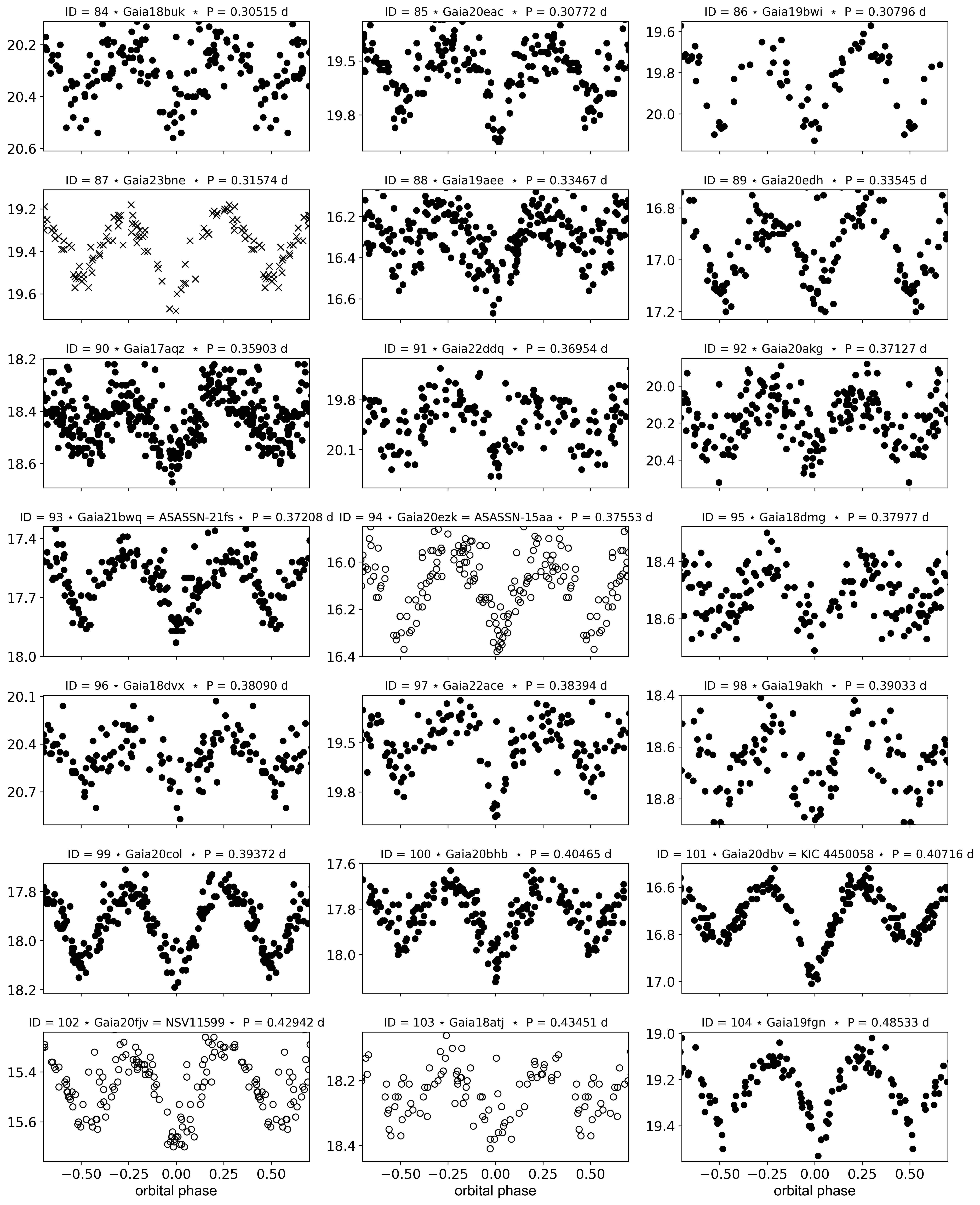}}
    \caption{\textit{– continued}}
    \label{fig5}
\end{figure}

\setcounter{figure}{0}
\begin{figure}
    \centerline{\includegraphics[width=18.5 cm]{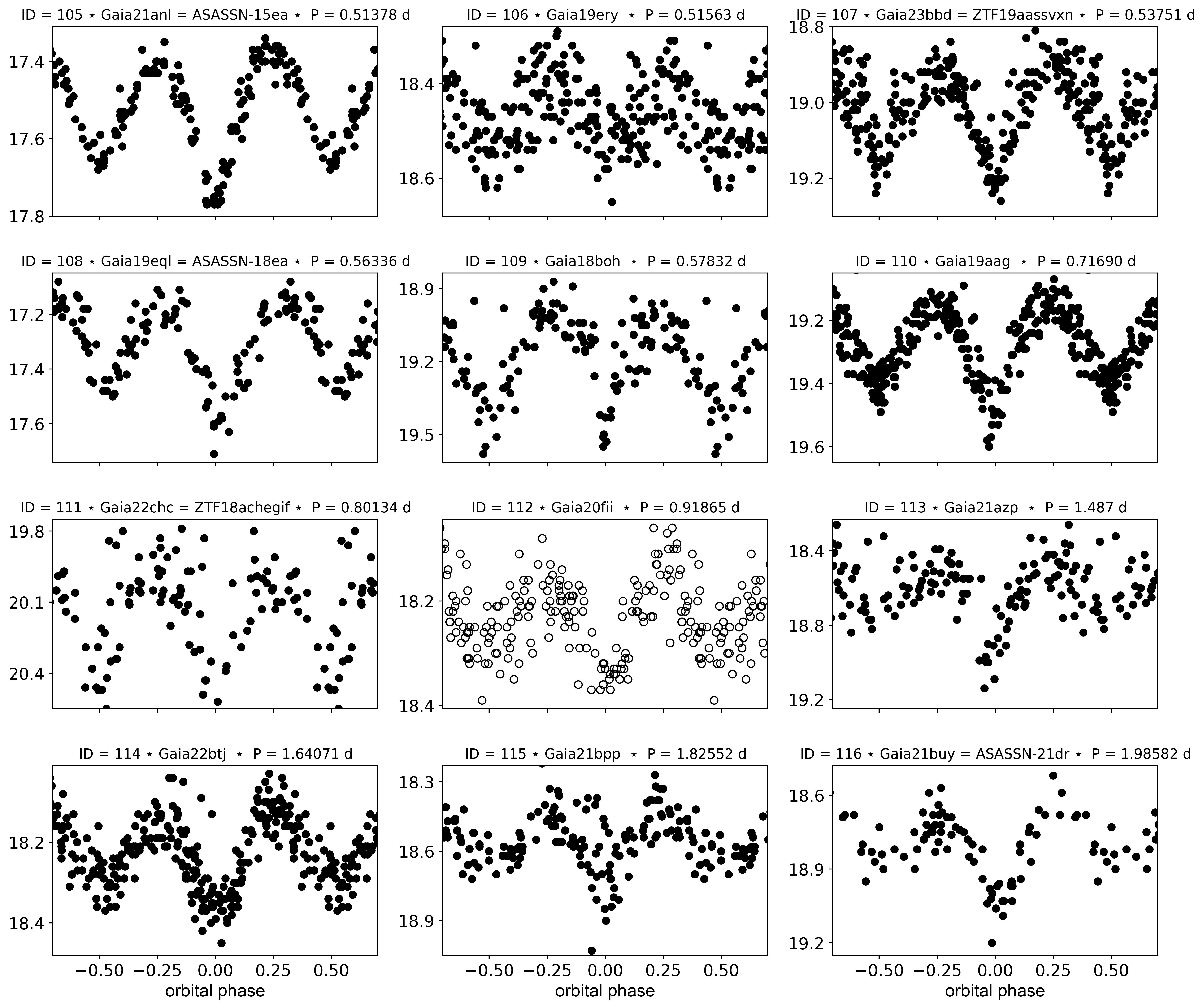}}
    \caption{\textit{– continued}}
    \label{fig6}
\end{figure}

\end{appendix}
\end{document}